\documentclass[aps,prl,amssymb,showpacs,amsmath,superscriptaddress,preprintnumbers,reprint]{revtex4-2}

\usepackage{graphicx}
\usepackage{dcolumn}
\usepackage[normalem]{ulem}

\usepackage{color}
\usepackage{xcolor}

\renewcommand{\epsilon}{\varepsilon}

\begin{document}

\title{Sequence of phase transitions in a model of interacting rods}

\author{Juliane U. Klamser}
\affiliation{
Gulliver UMR CNRS 7083, ESPCI Paris, Universit\'e PSL, 75005 Paris, France.}
\author{Tridib Sadhu}
\affiliation{Department of Theoretical Physics, Tata Institute of Fundamental Research, Mumbai 400005, India.}
\author{Deepak Dhar}
\affiliation{Department of Physics, Indian Institute of Science Research and Education, Pune 411008, India.}

\begin{abstract}
In a system of interacting thin rigid rods of equal length $2 \ell$ on a 
two-dimensional grid of lattice spacing $a$,  we show that there are multiple 
phase transitions as the coupling strength $\kappa=\ell/a$ and the temperature 
are varied.  There are essentially two classes of transitions.  One corresponds 
to the Ising-type spontaneous symmetry breaking transition and the second 
belongs to less-studied phase transitions of geometrical origin.  The latter class 
of transitions appear at fixed values of $\kappa$ irrespective of the 
temperature,  whereas the critical coupling for the spontaneous symmetry 
breaking transition depends on it.  By varying the temperature,  the phase 
boundaries may cross each other,  leading to a rich phase behaviour with 
infinitely many phases.  Our results are based on Monte Carlo simulations on the 
square lattice,  and a fixed-point analysis of a functional flow equation on a 
Bethe lattice.  
\end{abstract}

\pacs{05.40.Jc, 02.50.Cw, 87.10.Mn}

\maketitle
Lattice models with interacting degrees of freedom are 
extensively studied in statistical mechanics \cite{baxter2007exactly,friedli_velenik_2017}.   
They remain instructive 
textbook examples for emergent phases and phase transitions.  Especially 
in two dimensions,  the marginal 
competition between fluctuations and order can lead to interesting
phase behaviour.  Yet,  there are 
still surprises lurking even with the simplest kind of two-dimensional models.  
In this \textit{Letter} we 
present one such example where we find less-studied geometrical 
transitions coexisting alongside conventional disorder-order transitions.  An 
interplay between different mechanisms of these transitions  yields a rich 
phase diagram with an infinity of phases. These features are expected
to be not specific to the model studied here.

Our model is similar to the lattice-gases of rigid orientable molecules that were 
extensively studied for organic solids \cite{Huckaby1982,STAVELEY196146},  
hydrogen bonded solvents \cite{Abraham1972},  water \cite{Heilmann1979},  
and absorbed monolayers \cite{Southern_1979}.  In these models, anisotropic 
structure blocks (molecules) are pivoted at regular lattice sites,  but they can rotate while being constrained by 
steric interaction with neighbours.  This makes them different from well-known 
lattice-gases of movable molecules \cite{ONSAGER1949,Lebowitz1971,kantor2009,gurin2011}, 
particularly the hard-rod models for isotropic-nematic transitions without a positional 
order in liquid crystals \cite{Flory1956,Agren1974,HEILMANN1977,Ghosh_2007}.  
A counterpart of liquid crystals is plastic crystal phases 
\cite{TIMMERMANS19611,Folmer2008,Towle1973,Li2019} of organic solids,  where 
molecules are ordered in position but can show a varying degree of orientational disorder.   
Plastic crystals show a sequence of phase transitions from one crystalline phase to 
another,  when the temperature is lowered or the pressure is increased
\cite{Li2019,CHANDRA1991159,GUTHRIE196153,YASHONATH198522}.  Our simple 
model captures this basic phenomenology.

\begin{figure}[t]
\includegraphics{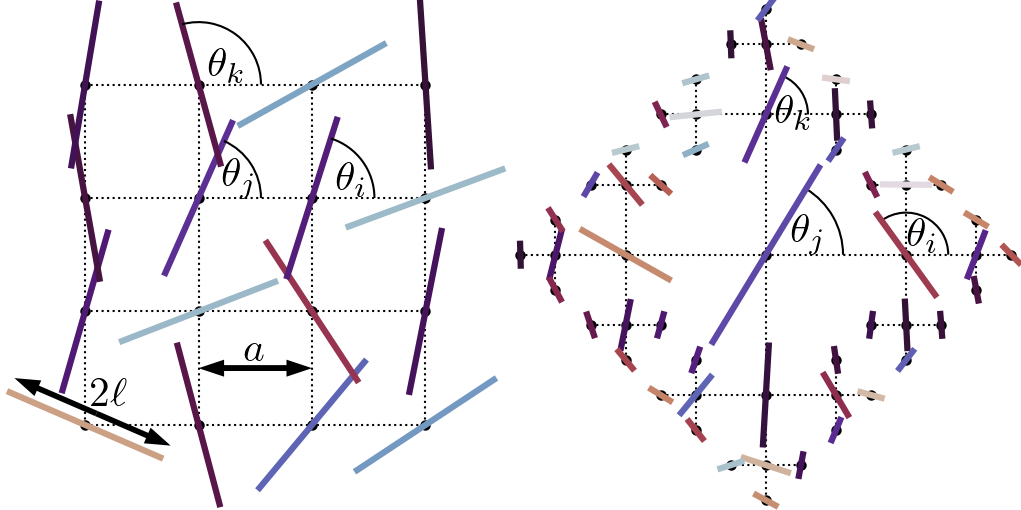}
\caption{(a) A configuration of rods on a square grid (dashed 
lines), where five overlaps 
between nearest-neighbour rods cost a total energy $ 5 U_1$.  (b) 
The model on a Bethe lattice (dashed lines), where, unlike the square lattice,  
only rods connected by a single lattice edge interact. All rods and lattice 
units have constant length, $2\ell$ and $a$  respectively. The varied scale in (b) 
 is for representation purpose only. \label{fig:rods}}
\end{figure}

In the simplest example we discuss in this \textit{Letter},  non-polar thin rigid rods of length 
$2\ell$ are pinned at their midpoint at the lattice sites of an $L\times L$ square grid of 
lattice unit $a$ and with periodic boundaries.  
Rods are free to rotate in the plane  as illustrated in 
Fig.~\ref{fig:rods}a.  The orientation of a rod at a site $\mathbf{r}$ is 
specified by its angle $\theta_{\mathbf{r}}$ with $0\le \theta_{\mathbf{r}} < 
\pi$.  A configuration is given by the set of angles of all rods.  
Interactions between rods are solely due to overlaps, 
 which costs a constant energy $U_1$ for each nearest neighbour overlap,  $U_2$ 
for each next nearest neighbour overlap, and so on, no matter where they occur. 
The total energy of a configuration $H=\sum_i n_i U_i$ with $n_i$ being the 
number of $i$-th neighbour overlaps.  Configurations are 
weighted by the canonical Boltzmann-Gibbs measure $e^{-\beta H}$ where $\beta$ is the 
inverse temperature.  (Similar models of orientable
anisotropic objects pivoted at lattice points have been studied in \cite{casey1969,Freasier1973,Deepak}.)

Using Markov-Chain Monte Carlo (MCMC) simulations we show that the system 
undergoes infinitely many phase transitions as $\beta$ and the coupling strength 
$\kappa=\ell/a$ are varied.  There are essentially two classes of transitions.  
One that belongs to the disorder-order transition,  where there is a breaking of 
the discrete symmetry group of the orientations in the model.  With increasing 
$\kappa$,  as more neighbours interact,  more symmetry breaking transitions take place, 
successively increasing the degree of orientational order.   The second class of transitions 
induce singularities in the angular distribution of rods $P(\theta)$.  These 
transitions are related to singular changes in the geometry of overlap manifold 
and we refer to them as geometrical transitions. They are induced by non-analyticities in 
the interaction potential between the rods as a function of the distance between them,
which, unlike the disorder-order transition, do not depend on $\beta$.   Each
geometrical and symmetry-breaking 
transition may be characterized by a different order parameter and an interplay 
between these transitions leads to infinitely many phases on the 
$(\beta,\kappa)$ plane.  

The model is generalizable to arbitrary lattices and higher dimensions.  In 
particular,  we discuss the model on a Bethe lattice 
shown in Fig.~\ref{fig:rods}b,  where the distribution of angular orientation 
is obtained from a fixed point solution of a functional equation 
corresponding to the self-consistency condition on the Bethe lattice.

\begin{figure}
\includegraphics{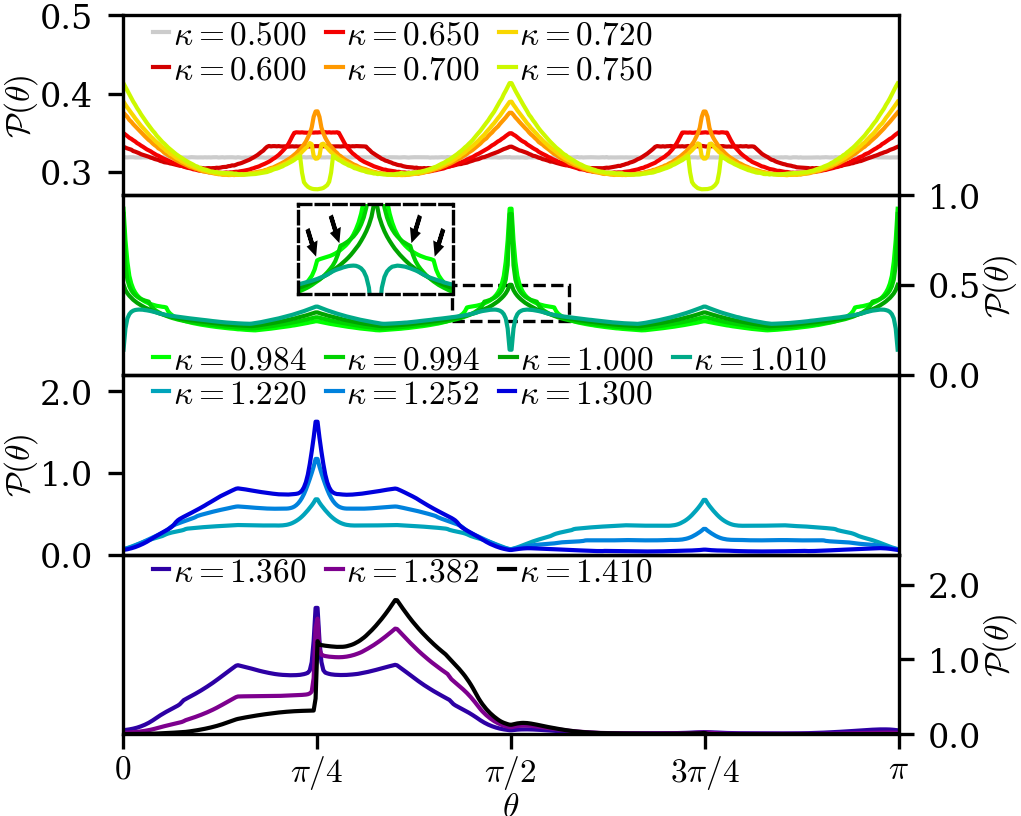}
\put(-50,185){(a) }
\put(-50,140){(b) }
\put(-50,95){(c) }
\put(-50,50){(d) }
\caption{Square lattice~-- (color online) Probability distribution of the 
orientation of rods at increasing values of $\kappa$.  (a): As $\kappa$ is 
increased from $0.5$ to $0.75$,  two pairs of derivative discontinuities cross 
each other at $\theta=\{\frac{\pi}{4},\frac{3\pi}{4}\}$ and become cusp 
singularities above $\kappa\simeq0.707$.    (b): The cusp singularities merge in 
pairs for $\kappa = 1$ at $\theta=\{0,\pi/2\}$.  Cusps are indicated by arrows 
in the zoom in the inset.  (c): The symmetry around $\theta = \pi/2$ still 
present for $\kappa = 1.22$,  is broken for $\kappa = 1.252$ and above.  (d): In 
the angular range $\theta \in (0,\pi/2)$ the symmetry around $\pi/4$ present for 
$\kappa=1.360$ is lost for $\kappa=1.382$ and above.  Results are for 
$\beta=1.4$ and $U_i = 1$ for all $i$ on an $L\times L$ square lattice with $L = 
240$.} \label{fig:derivative}
\end{figure}
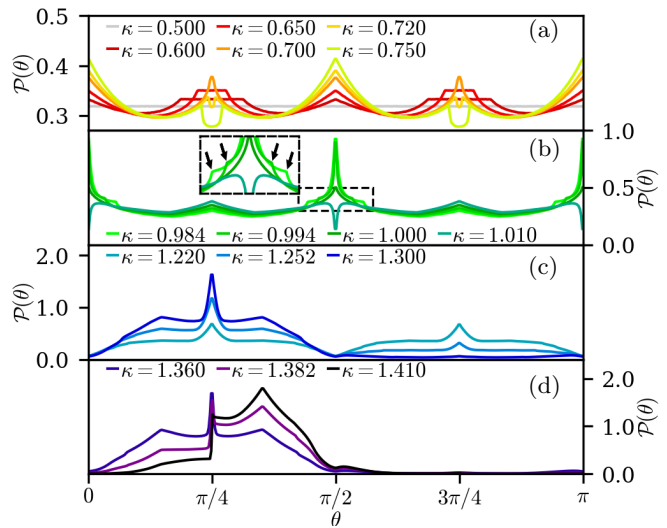

\textit{Probability distribution of orientations ---} For the model on a square lattice, 
first few transitions for small $\kappa$ are 
shown in the distribution $P(\theta)$ in 
Fig.~\ref{fig:derivative} for soft repulsive rods at $\beta = 1.4$, with $U_i = 1$ for all $i$.  
For $\kappa<1/2$,  rods do not interact and 
$P(\theta)$ is uniform.  For larger $\kappa$,  the distribution develops a 
non-trivial profile with $P(\theta)$ constant in some range,  and not in others, but has the four-fold 
symmetry of the lattice (see Fig.~\ref{fig:derivative}a).  The extent of the regions with constant
$P(\theta)$  decreases with increasing $\kappa$ and vanishes for $\kappa\simeq 0.707$.  As $\kappa$ 
crosses this value, the local maximum of $P(\theta)$ along the diagonal directions $\theta=\{\pi/4,3\pi/4\}$ 
becomes a local minimum,  and two pairs of cusp singularities emerge.  The location of these cusp singularities 
move with $\kappa$ and they are determined by the singularity in the overlap interaction.
The next sharp change happens as 
$\kappa$ crosses the value of one (see Fig.~\ref{fig:derivative}b). Firstly,  
two pairs of cusp singularities merge at  $\theta=\{0,\pi/2\}$ 
respectively.  Secondly,  the slope of $P(\theta)$ near these angles  
changes sign,  leading to a drop in $P(\theta)$.  A third singular change in 
$P(\theta)$ is at $\kappa \simeq 1.252$  where the symmetry around $\theta = 
\pi / 2$ is spontaneously broken (see Fig.~\ref{fig:derivative}c).  The probability $P(\theta)$ is depleted in 
one of the sectors $\theta\in(0,\pi/2)$ or 
$\theta\in(\pi/2,\pi)$, i.e. the majority of rods is preferentially 
aligned in either one of these sectors.  
In Fig.~\ref{fig:derivative}c,  we show 
the case where the first angular sector is spontaneously selected.
For $\kappa \gtrsim 1.382$, the reflection symmetry about $\theta=\pi/4$ 
is spontaneously broken (see Fig.~\ref{fig:derivative}d).

The first and the second singular changes in $P(\theta)$ 
(Fig.~\ref{fig:derivative}a and b) are about qualitative changes in the 
dependence of $P(\theta)$ on $\theta$ and correspond to the geometrical 
transitions.  For example, in the range $1/2<\kappa<1/\sqrt{2}$, $P(\theta)$ has 
flat parts, in the range $1/\sqrt{2} < \kappa < 1$, $P(\theta)$  has two pairs of 
square-root-cusp singularities, and so on.
The breakdown 
of symmetries of $P(\theta)$ (Fig.~\ref{fig:derivative}c and d) correspond to 
disorder-order transitions.  In the following we closely look at one from each 
class of transitions, namely, the geometrical transition at $\kappa=1$ and the 
spontaneous symmetry breaking transition at $\kappa\simeq1.252$.  

\begin{figure*}
\includegraphics[width=1\linewidth]{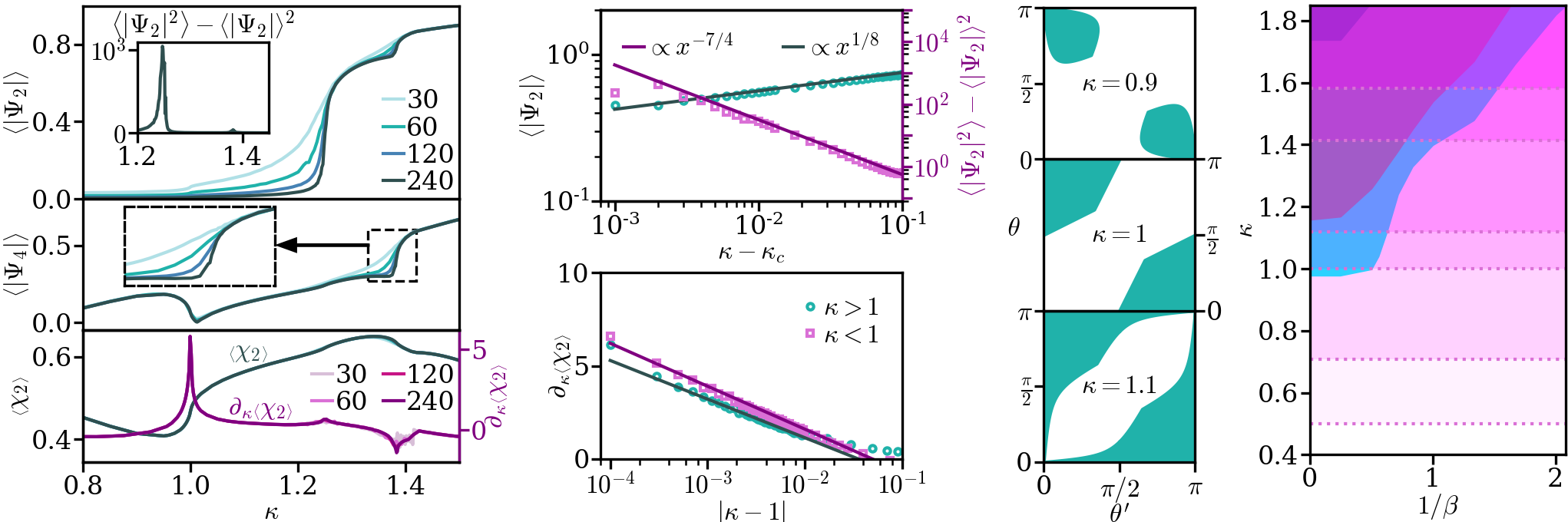}
\put(-483,110){(a) }
\put(-483,83){(b) }
\put(-483,41){(c) }
\put(-312,110){(d) }
\put(-312,26){(e) }
\put(-169,24){(f) }
\put(-83,26){(g) }
\caption{Square lattice~-- (color online) (a): Mean amplitude of the bulk 
orientation vector $\vert\Psi_2\vert$ for different system sizes $L\times L$ 
(legend indicates linear system size $L$) at $\beta = 1.4$ showing around $\kappa 
\simeq 1.252$ the typical 
finite-size behaviour of a continuous disorder-order transition.  
The variance of $\vert\Psi_2\vert$, shown in the inset for 
$L=240$, exhibits a sharp peak at the transition.  (b): Mean of order parameter 
$\vert\Psi_4\vert$ for the second disorder-order transition at $\kappa\simeq 
1.382$.  A finite size dependence is shown in the inset. (c): Mean of bulk order 
parameter $\chi_2$ shows an infinite slope at $\kappa= 1$,  which reflects as 
divergence in $\partial_\kappa \langle \chi_2\rangle$.  Shown data is for the 
same $\beta$ and system sizes as in (a),  but curves are almost indiscernible 
from each other due to absence of finite size effects around $\kappa=1$. (d): 
Power-law dependence of mean (round symbols) and variance (square symbols) of 
$\vert\Psi_2\vert$ for $\kappa\gtrsim \kappa_c$ for $L = 240$ and $\beta = 1.4$. 
 (e): Logarithmic divergence of $\partial_\kappa\langle \chi_2\rangle$ as 
$\kappa$ approaches one from both sides for $\beta = 1$ and $L = 60$.  (f): 
Overlap region,  filled with green color on the $(\theta,\theta')$ plane,  
between $x$-nearest-neighbour rods of orientation $\theta$ and $\theta'$.  
Boundary of the overlap region undergoes a sharp change as $\kappa$ crosses one. 
 (g): A sketch of the phase diagram where different coloured regions correspond 
to different phases.  Dotted horizontal lines denote geometrical transitions. 
The remaining phase boundaries correspond to disorder-order transitions. They 
are estimated from peak positions of the variance of the appropriate 
bulk-orientation order parameter at discrete points in the phase space.   
Results are for $U_i=1$ for all neighbours $i$.\label{fig:prob}}
\end{figure*}
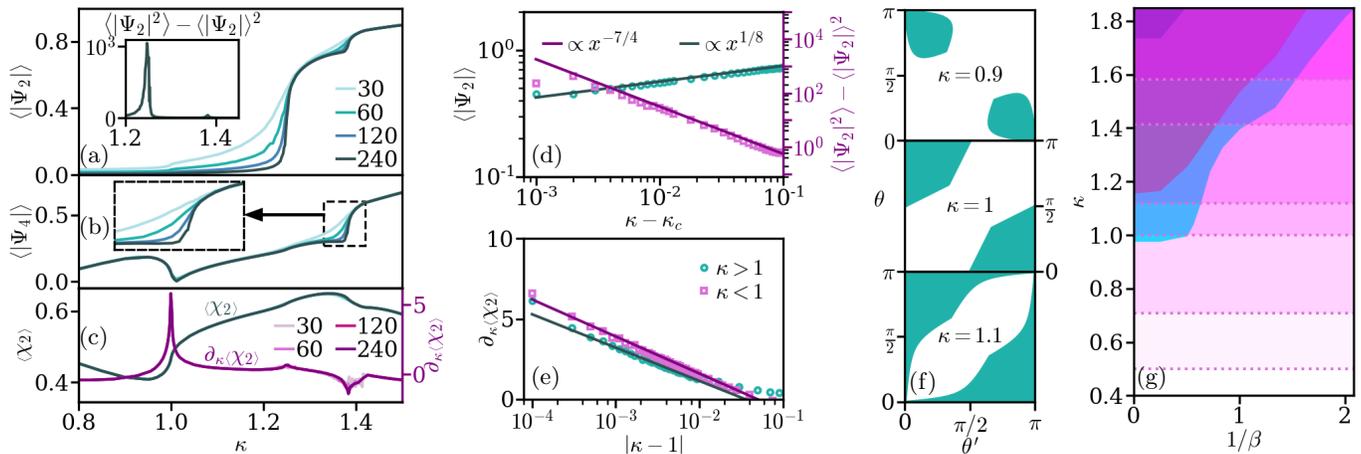

\textit{Symmetry breaking at $\kappa\simeq1.252$ ---} 
Considering the breakdown of $P(\theta)=P(\pi-\theta)$ symmetry in 
Fig.~\ref{fig:derivative}c, it is expected that the transition belongs to the 
Ising class of disorder-order transitions.  A measure of the long-range order 
set by the transition,  is the norm $\vert \Psi_2 \vert$ of the bulk orientation 
vector $\Psi_2=\tfrac{1}{L^2}\sum_{\mathbf{r}}\exp(i2\theta_{\mathbf{r}})$. (The 
observable is in analogy with the well-known nematic 
order parameter  $P_2(\cos\theta)$ in liquid crystals \cite{de1993physics} and the
bond-orientational order parameter  $\psi_6$ in the two-dimensional melting 
\cite{SteinhardtPsi6,KapferPsi6}.)   The change in average $\vert \Psi_2 \vert$ 
is shown in Fig.~\ref{fig:prob}a for various system sizes. 
Extrapolating the finite size effects to the thermodynamic limit gives $\vert \Psi_2\vert=0$ for $\kappa$ below a critical value 
$\kappa_c\simeq1.252$ and $\vert \Psi_2\vert$ non-zero 
above.

At an even higher value $\kappa\simeq1.382$,  above the onset of 
next-nearest-neighbour interactions,  $\vert \Psi_2 \vert$ shows the signature of 
a second symmetry-breaking transition.  Considering the broken symmetry in $P(\theta)$ in
Fig.~\ref{fig:derivative}d,  we consider $\vert \Psi_4\vert =\vert 
\tfrac{1}{L^2}\sum_{\mathbf{r}}\exp(i4\theta_{\mathbf{r}})\vert $ as an 
appropriate order parameter.  Fig.~\ref{fig:prob}b shows that while $\vert 
\Psi_4\vert$ is insensitive to the first disorder-order transition,  at the 
second transition it develops a singularity in the thermodynamic limit, with a behavior at 
finite $L$ consistent with the finite-size-scaling theory of continuous transitions. 
At increasing $\kappa$, there are more disorder-order 
transitions, induced by higher neighbour interactions,  each breaking finer 
symmetries. While the function $P(\theta)$ is expected to capture 
all these transitions,
in a Ginzburg-Landau-Wilson theory description, one would need different order 
parameters for different transitions.

\textit{Transition at $\kappa = 1$--} 
Considering the nature of changes in $P(\theta)$ in Fig.~\ref{fig:derivative}b, 
we construct a bulk observable $\chi_2=\tfrac{1}{L^2}\sum_\mathbf{r}(\sin 2\theta_\mathbf{r})^2$ 
that shows a strong variation at $\kappa=1$ shown in Fig.~\ref{fig:prob}c.  
We note two important features. First, there are no visible 
finite-size effects around $\kappa=1$ that could create an assymptotic 
singularity in the thermodynamic limit. Second, however, the slope of $\chi_2$ 
becomes infinite at $\kappa=1$ even for small systems, hence leading to the 
divergence of $\partial_\kappa\chi_2$, although the orientational correlation 
length remains finite. 

\textit{Universality class ---}
The divergence in $\partial_\kappa\chi_2$ at $\kappa = 1$ is in contrast to the discontinuous 
change in $\partial_\kappa\vert \Psi_2 \vert$ at $\kappa_c\simeq1.252$, consistent with the 
theoretical understanding of the different origins of the two transitions.  This 
difference is further evident 
from the scaling of these singularities.  For the disorder-order transition, 
the mean and the variance of 
$\vert\Psi_2\vert$ follow a power-law with vanishing $\kappa - \kappa_c$,  
\begin{equation}
\langle \vert \Psi_2 \vert\rangle \sim (\kappa - \kappa_{c})^{\beta}; \quad 
\langle \vert \Psi_2\vert^2\rangle -\langle \vert \Psi_2 \vert \rangle^2 \sim 
(\kappa - \kappa_{c})^{-\gamma} \,,\label{eq:power-law}
\end{equation}
(see Fig.~\ref{fig:prob}d) where the critical exponents $\beta=1/8$ and $\gamma=7/4$ 
are commensurate with 
the Ising exponents in two dimensions \cite{baxter2007exactly}.
In comparison, $\partial_\kappa \langle \chi_2\rangle$ diverges logarithmically 
with $\vert \kappa -1\vert$ (see Fig.~\ref{fig:prob}e).

\textit{Geometric origin of logarithmic divergence ---}
The unusual logarithmic singularity at $\kappa=1$ has a subtle geometric origin. 
 It relates to a non-analytic dependence of the overlap area in the 
$(\theta$,$\theta')$-plain of nearest neighbour rods as a function of $\kappa$.  
A pictorial representation of the overlap region between two 
$x-$nearest-neighbour rods of angles $\theta$ and $\theta'$ is shown in 
Fig.~\ref{fig:prob}f.  (Overlap regions for neighbours along the $y$-direction 
is related by symmetry.) The equation for the boundary of the overlap region is 
obtained from simple geometry.  For $\kappa$ near one,  the difference of 
the overlap area from its value at $\kappa=1$ vanishes as $\epsilon=\vert 
\kappa-1\vert $ tends to zero,  but only as $\epsilon \log \epsilon$.  This 
logarithmic dependence leads to the logarithmic divergence of $\partial_\kappa 
\chi_2$ in Fig.~\ref{fig:prob}e.  

The geometrical mechanism behind the phase transition was first reported 
\cite{SaryalRods1D} in one-dimension using a transfer-matrix method. (For a 
similar origin of singularities in pair-correlation function of 2-d disks see 
\cite{Stillinger}.) An exact solution in higher dimensions  has been possible
only for a small range of $\kappa$ \cite{Deepak}.
Nevertheless, the geometrical 
transition at $\kappa = 1$ can be established from its characteristics, notably 
the logarithmic divergence. Moreover, unlike in a conventional phase transition, 
non-analyticities appear even for small systems.  
Indeed,  we observe a negligable system size dependence around the singularity 
at $\kappa = 1$ in Fig.~\ref{fig:prob}c.  Another signature 
is that the critical value of $\kappa$ does not depend on $\beta$ as 
long as it remains non-zero. 

There are additional geometrical transitions from the overlap between 
second-nearest-neighbour rods,  third-nearest-neighbour rods, and so on.  In 
fact, the overlap function for any pair of rods depends only on the distance between 
the two centres, and is exactly the same as in the model in one dimension
 studied earlier \cite{SaryalRods1D}, which also shows a logarithmic singularity 
 at $\kappa = 1$. 
 The overlap function for nearest neighbours undergoes additional 
 singular changes at $\kappa = 1/2$ (onset of overlap interactions) 
 and $\kappa = 1/\sqrt{2}$ (see \cite{SaryalRods1D} for details), which reflect in $P(\theta)$
in Fig.~\ref{fig:derivative}a.
Therefore, including larger neighberhood interactions on the square lattice, where sites 
 are indexed by pairs of integers $(m,n)$, geometrical transitions happen at 
every $\kappa=z\sqrt{m^2+n^2}$,  for $z=\{1/2,1/\sqrt{2},1\}$.  (While $z=1$ 
is redundant, it is kept to emphasize the relation of $z$ to the different origins of the
geometrical singularities in the overlap function when the rod length
is half,  $1/\sqrt{2}$, or equal to the distance between the rod-centres.)
 
\textit{Phase diagram ---} Tracing the phase boundaries of geometrical 
and symmetry-breaking transitions in the $(\beta,\kappa)$ plane gives
a rich phase diagram shown in Fig.~\ref{fig:prob}g.  
Unlike the geometrical transitions,  the critical $\kappa$ for the disorder-order 
transitions depends on $\beta$.  The infinite $\beta$ limit corresponds to 
hard-rod interactions where no overlap is allowed.  In this limit the first 
disorder-order transition is at $\kappa_c \simeq 0.983$, and in this case 
the transition is entropy driven similar to the symmetry breaking transition 
in hard-spheres \cite{AsakuraOosawa,Leibler,krauth2006statistical}. 

Establishing the phase diagram analytically, even for the simplest case where 
only nearest neighbours interact ($U_i=0$ for $i>1$), requires solving for the 
largest eigenvalue of a multi-variable transfer operator analogous to the column 
transfer matrix for the two-dimensional Ising model 
\cite{SchultzMattisLiebIsing}.   An instructive, more tractable case is 
the Bethe lattice which we discuss below.  

\textit{Bethe lattice ---} As an analogue of the square lattice we consider a 
four-coordinated Bethe lattice shown in Fig.~\ref{fig:rods}b, where only rods 
linked by a single lattice edge interact, 
i.e. $U_i=0$ for $i>1$.  Since overlaps depend on the spatial orientation of the 
rods, we assign to each lattice edge a label x or y depending on the edge 
orientation in the plane. By construction, the overlap interaction between 
nearest neighbour rods is the same as on the square lattice.

A Bethe lattice is a tree,  thus devoid of loops.  It is therefore possible to 
write a recursion relation for branches at different depths, which led to exact 
solutions of well-known models 
\cite{baxter2007exactly,Dhar_1990_SandPile,Peruggi_1983}.  Adapting this idea 
for our model on the Bethe lattice, we obtain \cite{KSD2} the probability 
distribution in the thermodynamic limit,

\begin{equation}
P(\theta)= 
\frac{1}{Z}\left[g^\star_x(\theta)g^\star_y(\theta)\right]^2,\label{eq:angle 
distribution thermo}
\end{equation}
where $Z$ is the normalization,  and $(g^\star_{x},g^\star_{y})$ is the stable 
fixed point solution 
\begin{equation}
R_{x}(g_x^\star,g_y^\star)=g_x^\star, \qquad R_y(g_x^\star,g_y^\star)=g_y^\star 
\label{eq:fixedpoint}
\end{equation}
of the functional flow equations $R_{x}(g_x,g_y)\to g_{x}$ and $R_{y}(g_x,g_y)\to g_{y}$ with 
an initial condition $g_{x}(\theta) = g_{y}(\theta) =1$ and the generator of the flow 
\begin{equation}
R_{x}(g_x,g_y)[\theta] =  \mathcal{N}\cdot  \int_0^\pi \frac{d\theta'}{\pi} 
T_{x}(\theta,\theta')g_{x}(\theta') g_{y}^2(\theta'),
\label{eq:rec reln x}
\end{equation}
($R_{y}$ analogously) where the operator $\mathcal{N}\cdot f(\theta)$ 
resets the maximum of 
$f(\theta)$ to one,  $T_{x}(\theta,\theta')=e^{-\beta U_1}$ when a pair of 
$x-$neighbour rods of angles $\{\theta,\theta'\}$ overlap and is equal to one 
otherwise.  The $T_x$ is constructed from the overlap region in 
Fig.~\ref{fig:prob}f,  and $T_y$ from the former using symmetry. 

For $\kappa<\frac{1}{2}$, where rods are not interacting and 
$T_{x}(\theta,\theta')= T_{y}(\theta,\theta') =1$, we see 
$g_{x}^\star(\theta) = g_{y}^\star(\theta)=1$ is a stable 
fixed point,  which gives a uniform distribution $P(\theta)$ in 
Eq.~\eqref{eq:angle distribution thermo}.  For $\kappa>\frac{1}{2}$,  the fixed 
point is hard to solve analytically.  Nevertheless,  a numerical solution is 
straightforward by implementing the flow equation with a discretised angular 
range.

\begin{figure}
\includegraphics{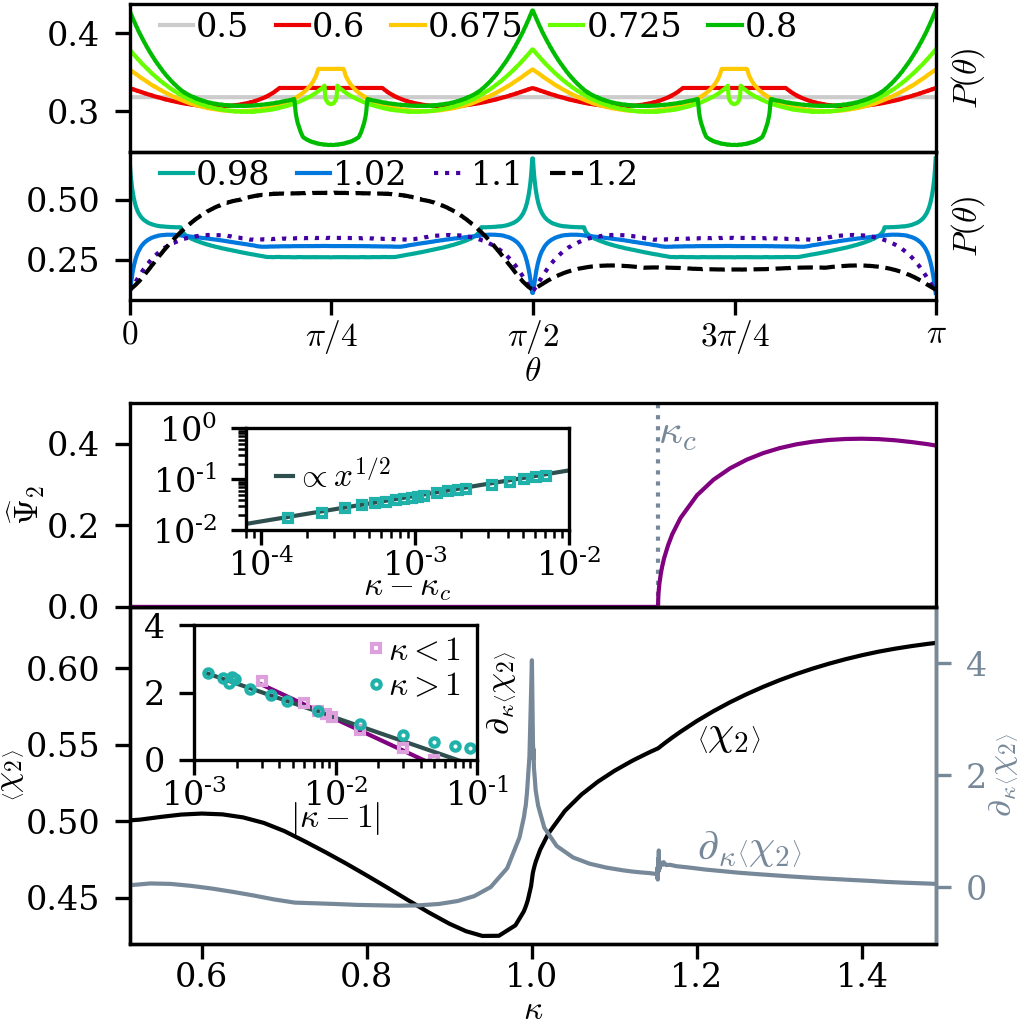}
\put(-34,213){(a) }
\put(-34,199){(b) }
\put(-34,130){(c) }
\put(-34,75){(d) }
\caption{Bethe lattice ~-- (color online) (a) and (b): Distribution of orientations $P(\theta)$ at 
various $\kappa$ indicated in legend. (a): $P(\theta)$ is non-uniform 
for $\kappa > 1/2$, with flat parts that vanish at $\kappa = 1/\sqrt{2}$, above which
$P(\theta)$ develops two pairs of square-root cusp singularities, that move apart from
each other. (b): At $\kappa = 1$ cusp singularities merge at $\theta = \{0, \pi/2\}$, 
where $\partial_\kappa P(\theta)$ also changes sign. 
While $P(\theta)$ is still symmetric around
$\pi/2$ for $\kappa = 1.1$ (dotted curve), this symmetry is broken for $\kappa = 1.2$ 
(dashed curve).
(c): Continuous disorder-order transition at 
$\kappa_{c}\simeq 1.1529$ seen in the order parameter $\widehat{\Psi}_2$, which 
vanishes as a power-law (see inset) with decreasing $\kappa - \kappa_c \ge 0$. 
(d): Infinite slope of $\langle\chi_2\rangle$ and diverging 
$\partial_\kappa\langle\chi_2\rangle$ at the geometrical transition point 
$\kappa = 1$. The inset shows characteristic logarithmic divergence of  
$\partial_\kappa\langle\chi_2\rangle$ as the transition point is approached. 
Data for $\beta U_1=1.0$ and an angular discretisation with $\Delta \theta = 
\pi/1200$.  }
\label{fig:order parameter}
\end{figure}
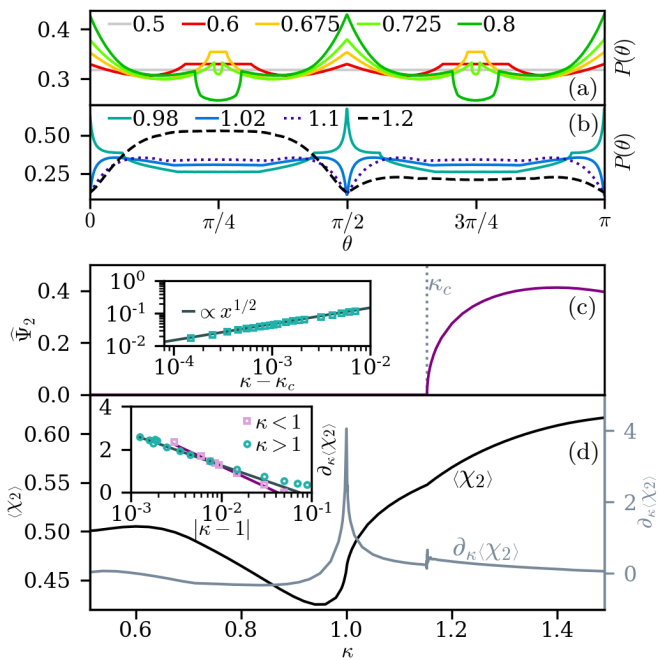

Our numerical iteration shows that there is a critical value $\kappa_{c}$ that 
depends on $\beta$, below which there is only one stable fixed point. The 
corresponding $g_{x}^\star(\theta)$ and $g_{y}^\star(\theta)$ are not uniform, but they are 
commensurate with the symmetry around $\pi/2$. The $P(\theta)$  
from Eq.~\eqref{eq:angle distribution thermo} shows qualitative agreement with 
the square-lattice results (compare Fig.~\ref{fig:derivative} with Fig.~\ref{fig:order parameter}a-b). 
For $\kappa>\kappa_{c}$,  the symmetric fixed point becomes unstable and two new 
stable fixed points appear which break the symmetry around $\pi/2$.  
The recursion flows into either of the two fixed points with equal probability, 
corresponding to two equally probable states with broken symmetry, of which one is
spontaneously selected in the thermodynamic limit. 

The bulk order parameter $\Psi_2$ involves spatial correlations, which requires
the joint probability $P(\theta,\theta')$. We chose an analogous 
quantity $\widehat{\Psi}_2 =| \langle e^{i2\theta}\rangle |$
where $\langle \cdot \rangle$ denotes averages with $P(\theta)$,  
that captures the continuous disorder-order transition as shown in 
Fig.~\ref{fig:order parameter}c (compare with Fig.~\ref{fig:prob}a).  Near the transition,  the 
power-law $\widehat{\Psi}_2 \sim ( \kappa - \kappa_c)^{\beta}$ with 
the critical exponent $\beta=1/2$ corresponds to the mean field Ising 
universality \cite{baxter2007exactly}. 

The geometrical transition is shown in Fig.~\ref{fig:order parameter}d.  Similar 
to the square lattice, the transition point $\kappa=1$ is unaffected by changes 
in $\beta$ and the bulk observable $\langle \chi_2 \rangle $ shows a 
characteristic logarithmic singularity in its derivative. 
There are two additional geometrical transitions at $\kappa = \{1/2, 1/\sqrt{2}\}$.
The relation of these transitions to the singular changes in the overlap region 
(see e.g. Fig.~\ref{fig:prob}f) can be seen from Eqs.~ 
\eqref{eq:angle distribution thermo}-\eqref{eq:rec reln x}.

To conclude, several comments are in order. 
Unlike the square lattice where additional 
transitions are induced by higher-neighbour 
overlaps (see Fig.~\ref{fig:prob}g ), the Bethe lattice has by construction no 
additional transitions.  However, for the range in 
$\kappa$, where both lattices can be compared,  the
qualitative behaviour of the orientation distribution $P(\theta)$ as a
function of $\kappa$ and $\beta$ is fully captured in this approximation.
A similar qualitative phase behaviour with multiple transitions can 
be seen \cite{KSD2} for the triangular lattice and it is expected to be 
robust for different shapes of molecules,  lattices,  and in 
higher dimensions. 

Note that we can think of the full function $P(\theta)$ as 
the order parameter, or equivalently an infinity of order parameters,
somewhat like the overlap function in the replica theory of spin glasses \cite{Mezard}.
The infinity of phases is also reminiscent of ANNNI type models \cite{SELKE1988213}.

A comparison of critical exponents shows that on the square lattice
the symmetry breaking transitions are Ising-type.  However, they can be different for other lattice geometries,  like a triangular lattice.  Note that the fixed lattice directions break the continuous symmetry,  and the Mermin-Wagner theorem does not apply.

The geometrical transitions are different from the familiar 
phase transitions in the Ehrenfest classification.  Corresponding order 
parameters change continuously like in a continuous transition,  however 
correlation lengths remain finite \cite{SaryalRods1D,KSD2}.  The 
geometrical transitions are expected to be generic to models where 
interactions have a non-analytic dependence on control parameters, e.g.  
an Ising model with an exchange interaction $J(r)$ that is a 
non-analytic function of the distance $r$.  

The question remains to what extent the many transitions seen in our model 
realize in natural examples.  There are known examples of multiple ordered states 
in organic materials, e.g.  Methane absorbed graphite is a  
two-dimensional example that undergoes an orientational ordering transition \cite{Huckaby1982}.  
In three dimensions,  plastic crystals like Neopentylglycol  \cite{Li2019} 
order in their molecular orientation by cooling below a certain temperature or from 
lattice contraction by hydrostatic pressure.   These changes in orientational order are
seen in diffraction experiments \cite{Temleitner_2013,More_1980}. 
It would be interesting to see if the model discussed in this \textit{Letter}, 
possibly with additional non-hard-core couplings, can be used to make quantitative agreement 
with the transitions in experimental systems.   The large caloric effect of ordering 
transitions in plastic crystals makes these questions practically relevant \cite{Li2019,CHANDRA1991159}.

\acknowledgements
We thank Sushant Saryal for stimulating discussions.
The work of DD is supported by a Senior Scientist Fellowship from 
the National Academy of Sciences of India. TS acknowledges support 
of the Department of Atomic Energy, Government
of India, under Project Identification No. RTI-4002.  TS acknowledges 
hospitality of the Gulliver Laboratory at ESPCI Paris $\vert$ PSL, where 
part of the work was completed.

\bibliography{citation}

\end{document}